# Biomass water content effect on soil moisture assessment via proximal gamma-ray spectroscopy

Marica Baldoncini[a,b,*], Matteo Albéri[a,b], Carlo Bottardi[b,c], Enrico Chiarelli[a,b], Kassandra Giulia Cristina Raptis[a,b], Virginia Strati[b,c], Fabio Mantovani[b,c]

[a]INFN, Legnaro National Laboratories, Viale dell'Università, 2, 35020, Legnaro, Padua, Italy

[b]Department of Physics and Earth Sciences, University of Ferrara, Via Saragat 1, 44121, Ferrara, Italy

[c]INFN, Ferrara Section, Via Saragat 1, 44121, Ferrara, Italy

*Corresponding author.

E-mail address: baldoncini@fe.infn.it (Marica Baldoncini)

**Abstract**

Proximal gamma-ray spectroscopy supported by adequate calibration and correction for growing biomass is an effective field scale technique for a continuous monitoring of top soil water content dynamics to be potentially employed as a decision support tool for automatic irrigation scheduling. This study demonstrates that this approach has the potential to be one of the best space–time trade-off methods, representing a joining link between punctual and satellite fields of view. The inverse proportionality between soil moisture and gamma signal is theoretically derived taking into account a non-constant correction due to the presence of growing vegetation beneath the detector position. The gamma signal attenuation due to biomass is modelled with a Monte Carlo-based approach in terms of an equivalent water layer which thickness varies in time as the crop evolves during its life-cycle. The reliability and effectiveness of this approach is proved through a 7 months continuous acquisition of terrestrial gamma radiation in a 0.4 hectares tomato (Solanum lycopersicum) test field. We demonstrate that a permanent gamma station installed at an agricultural field can reliably probe the water content of the top soil only if systematic effects due to the biomass shielding are properly accounted for. Biomass corrected experimental values of soil water content inferred from radiometric measurements are compared with gravimetric data acquired under different soil moisture levels, resulting in an average percentage relative discrepancy of about 3% in bare soil condition and of 4% during the vegetated period. The temporal evolution of corrected soil water content values exhibits a dynamic range coherent with the soil hydraulic properties in terms of wilting point, field capacity and saturation.

## 1. Introduction

Soil water content (SWC) is a relevant state variable tracking the exchange of water at the land surface and is a key to understand and predict soil hydrological processes over a broad range of scales (Vereecken et al., 2015). Tracing its dynamics provides essential information for a deeper understanding of the major hydrological, biogeochemical, and energy exchange processes (Brocca et al., 2017), as well as for improving water use efficiency in agriculture, which is definitely the main competitor in the worldwide race to water resources (Levidow et al., 2014; Ozbahce and Tari, 2010). Therefore, technological and methodological advancements are highly desired for accurate measurements of the spatial and temporal SWC variability (Michot et al., 2003; Robinet et al., 2018; Sultana et al., 2017).

Recently, proximal and on-the-go soil sensors are being widely adopted for understanding soil properties and hydrogeological processes in precision agriculture (Heggemann et al., 2017; Piikki et al., 2015; Viscarra Rossel et al., 2007). From one side they have a relatively wider spatial coverage compared to point scale sensors, and from the other side they are less subject to interfering factors (e.g. atmospheric effects or observation conditions in terms of intensity and direction of illumination) in comparison to traditional remote sensing methods based on satellite spectral images (Barnes et al., 2003; McBratney et al., 2003). In this scenario, permanently installed measurement stations for proximal gamma-ray spectroscopy match the current requirements for SWC sensing methods as they (i) keep the soil structure undisturbed during the data taking, (ii) operate continuously allowing for a characterization of the SWC temporal dynamics and (iii) integrate measurements at the field scale over areas of 1 to about 10 km$^2$ (Bogena et al., 2015; Strati et al., 2018).

Gamma-rays are high-energy photons continuously produced in soils due to the presence of $^{40}$K and daughter products of the $^{238}$U and $^{232}$Th decay chains. As the signal recorded by a spectrometer provides clues on the propagation of gamma-rays from the emission to the detection point, environmental gamma spectra probe at the same time the activity of the radioactive source and the physical-chemical properties of the traversed materials in terms of different attenuation effects, the latter essentially dominated by material density and consequently by SWC (Minty, 1997).

Environmental gamma-ray spectroscopy measurements are influenced by plenty of experimental boundary conditions which knowledge help in interpreting radiometric data at different levels according also to the spatial scale of the surveys. Airborne gamma-ray spectroscopy already raised the attention on the attenuating effects on the gamma signal due to the presence of vegetation (Dierke and Werban, 2013; Norwine et al., 1979; Sanderson et al., 2004; Wilford et al., 1997). However, the presence of biomass in terms of plants, leaves and fruits is expected to play a much more critical role in proximal gamma-ray surveys, which implies that an accurate estimate of the signal reduction is needed.

The physical-chemical properties and the radioactive content of agricultural soils can be considered almost stationary, or at least sufficiently under control. The same does not apply to the crop system which is subject to a highly dynamic development generally affected by variable climatic conditions and irrigation management practices. Indeed, the presence of growing vegetation introduces a sizable extra attenuation due to the Biomass Water Content (BWC). The BWC varies in time during the crop life-cycle and causes a gamma-ray attenuation which is in principle undistinguishable from that generated by an increase in SWC. In this perspective, a reliable correction for the BWC shielding is mandatory in order to avoid a systematic overestimation of SWC.

The goal of this paper is evaluating the BWC attenuation effect in the framework of a proximal gamma-ray spectroscopy experiment performed at a tomato test field. The experiment was conducted by installing a permanent gamma station constituted by a 1L sodium iodide (NaI) detector placed at a height of 2.25 m, which collected gamma-rays emitted within an area of about 25 m radial distance and within a depth of approximately 30 cm. An ad-hoc gravimetric calibration campaign was performed by collecting soil and biomass samples. Experimental daily values of the SWC were estimated over a data taking period that lasted for 7 months and were evaluated by taking into account the shielding effect due to the presence of growing BWC during the tomato crop season.

## 2. Material and methods

In the following section we briefly present a geographical and climatic setting of the experimental site and a characterization of the main physical and hydraulic properties of the soil. The gamma and agro-meteorological stations are described together with the data acquisition methods. The gravimetric sampling campaign performed on soil and biomass samples is described along with the obtained results.

### 2.1. Experimental site

The experiment was conducted in the period 4[th] of April – 2[nd] of November 2017 at a tomato field of the Acqua Campus, a research center of the Emiliano-Romagnolo Canal (CER) irrigation district in the Emilia Romagna region in Italy (44.57° N, 11.53° E; 16 m above sea level) (Figure 1). According to the Köppen-Geiger climate classification (Peel et al., 2007), this geographical area is classified as Cfa (i.e. temperate, without dry season and with hot summer); its average annual temperature is 14 °C and rainfall is 700 mm.

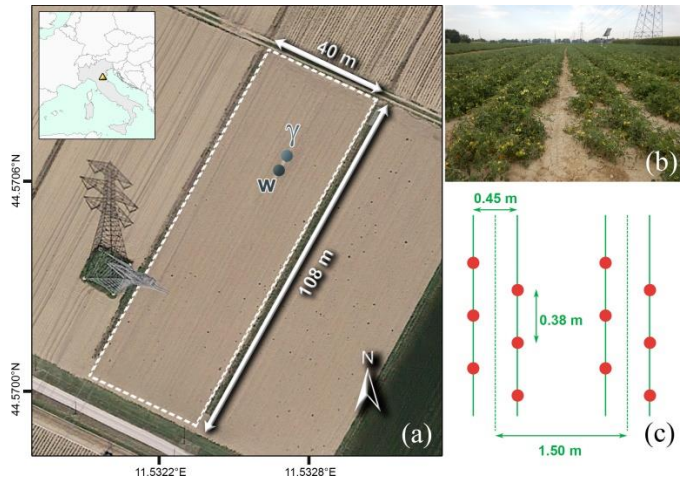

Figure 1. Panel (a), geographic location of the experimental site; the field dimensions and the positions of the gamma (γ) and agro-meteorological (w) stations are also reported (cartographic reference system WGS 84). Panel (b), picture of the tomato plants rows. Panel (c), schematic diagram of the disposition of the tomato plants rows.

About 24% of the agricultural territory in Emilia Romagna, one of the richest regions of Italy and Europe, is devoted to irrigated agriculture, which plays a major role in the regional economy (Munaretto and Battilani, 2014). In particular, Emilia Romagna is the Italian region having the largest surface of land cultivated with tomatoes, one of the most water-demanding crops among vegetables, and contributes for about one third of the tomato national production (ISTAT, 2017).

The main physical and hydraulic parameters of the soil, characterized by a loamy texture and a 1.26% organic matter content, are listed in Table 1 (after (Strati et al., 2018)).

Table 1. Physical and hydraulic parameters of the experimental site soil for the depth horizon [0–30] cm, after (Strati et al., 2018). Sand, silt, and clay percentages as well as bulk density and organic matter were determined from direct measurements. The wilting point, field capacity and saturation value were inferred from the water retention curve.

| Parameter | Value |
|---|---|
| Sand [%] | 45 |
| Silt [%] | 40 |
| Clay [%] | 15 |
| Soil textural class | Loam |
| Soil bulk density [kg/m$^3$] | 1345 |
| Wilting point [kg/kg] | 0.07 |
| Field capacity [kg/kg] | 0.24 |
| Saturation [kg/kg] | 0.36 |

Tomato plants were transplanted on the 23$^{rd}$ of May with a row and plant spacing as shown in Figure 1, which corresponds to a 3.5 plants/m$^2$ density, and harvested on the 14$^{th}$ of September. The crop phenological growth stages of anthesis and maturity, together with the planting and harvesting dates, are indicated in panel (a) of Figure 3. Irrigation water was delivered by a sprinkler system, following a schedule based on the criteria provided by the decision support tool of IRRINET (Munaretto and Battilani, 2014).

## 2.2. Experimental setup

The experimental setup is composed of a gamma spectroscopy station and a commercial agro-meteorological station (MeteoSense 2.0, Netsens) both powered by solar panels and provided with an internet connection (Figure 2).

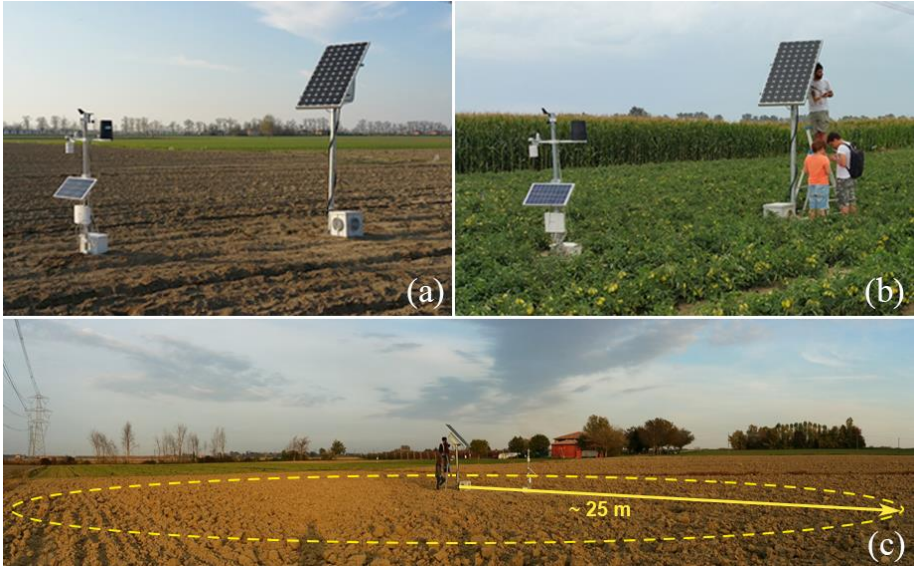

Figure 2. Panel (a) and (b) show the gamma and weather stations installed at the experimental site respectively in bare soil condition and during the vegetated period. Panel (c) illustrates a schematic representation of the gamma-ray spectrometer footprint at 2.25 m height: 95% of the detected signal comes from an area having radius of ~25 m.

The gamma station was specifically designed and built for the purpose of this experiment: its external structure is made up of steel and comprises a steel box welded on top of a 2.25 m high pole which hosts a 1L sodium iodide (NaI(Tl)) gamma-ray spectrometer (Baldoncini et al., 2018). The crystal is coupled to a photo-multiplier tube base which output is processed by a digital Multi-Channel Analyzer (MCA, CAEN γstream) having 2048 acquisition channels. At a height of 2.25 m about 95% of the detected gamma signal is produced within a cone having base radius of approximately 25 meters (Feng et al., 2009) (Figure 2).

The MCA is complemented with a small integrated computer which provides the necessary hardware interface to the detector and runs the software required for managing the acquisition parameters, namely the start time, the acquisition dynamics in terms of spectral gain [keV/ch], and the operating high voltage. Additional software was developed to make the data-taking continuous and more resilient to some hardware related failures like accidental restarts or power shortages.

Measured weather data include air temperature, relative air humidity, wind direction and speed, precipitation and Short Wave Incoming Radiation (SWIR). Figure 3 shows the daily values of Minimum and Maximum Temperatures ($T_{min}$ and $T_{max}$), ranging in the $T_{min}$ = [1.3 - 22.7] °C and $T_{max}$ = [13.5 - 39.3] °C intervals (panel a), the SWIR (ranging from 34.7 to 257.3 W/m$^2$) (panel b), the daily rainfall amount (up to a maximum of 56.2 mm) and irrigation water (up to a maximum of 35 mm) (panel c). The evapotranspiration (ET0, panel b) is calculated with the Hargreaves method (Hargreaves and Samani, 1985) using weather data recorded by the agro-meteorological station. During the last ten years (2008 - 2017) local meteorological archives (Arpae) recorded a mean total rainfall in the same period of 384.3 mm, a mean daily minimum temperature of 13.2 °C and a mean daily maximum temperature of 26.3 °C.

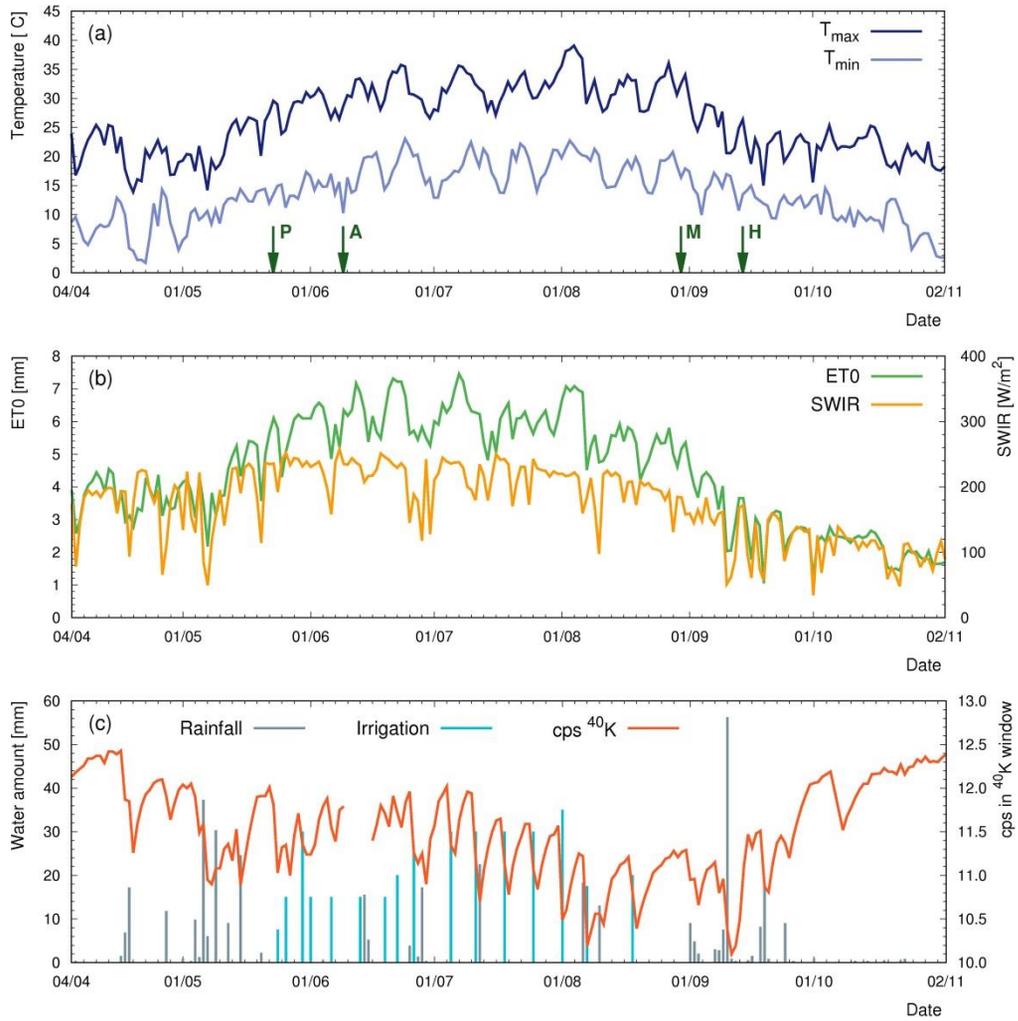

Figure 3. Daily meteorological and gamma data. In panel (a), maximum ($T_{max}$) and minimum ($T_{min}$) temperatures; the arrows indicate the crop stages of planting (P), anthesis (A), maturity (M) and harvesting (H). In panel (b), reference crop evapotranspiration (ET0) and Short Wave Incoming Radiation (SWIR). In panel (c), amount of rainfall and irrigation water and daily average counts per second (cps) in the $^{40}$K gamma photopeak energy window.

## 2.3. Data acquisitions

### 2.3.1. Gravimetric measurements

Gravimetric measurements were carried out on bulk soil samples as means to both calibrate and validate the soil water content estimation based on proximal gamma-ray spectroscopy. Five sets of samples to be characterized via gravimetric measurements were collected: (i) a calibration set collected in bare soil condition on the 18[th] of September one day before a rainfall event, (ii) a validation set collected in bare soil condition on the 21[st] of September two days after a rainfall event, (iii) three validation sets collected in presence of the tomato crop and one Day Before Irrigation (DBI) (24[th] of July), one (26[th] of July) and three (28[th] of July) Days After Irrigation (DAI) (Table 2).

Samples were collected by using a soil-auger following a sampling scheme (Figure 4) including 16 planar sampling points and covering homogeneously the area within a ~15 m radius from the gamma station position from which about 85% of the detected signal is produced (Figure 2 of (Baldoncini et al., 2018)).

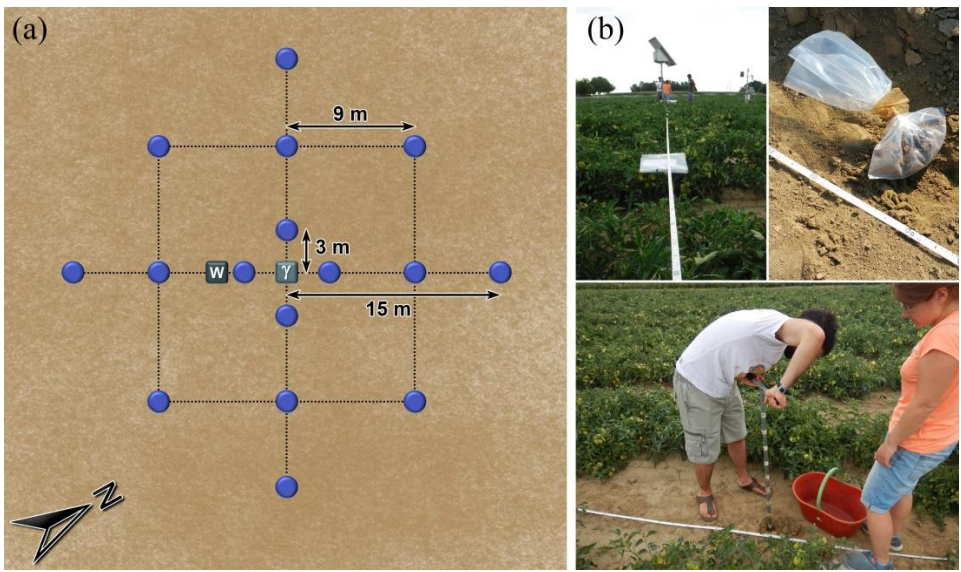

Figure 4. Panel (a), scheme of the 16 collection points adopted for the soil gravimetric sampling campaign together with their relative distances to the gamma ($\gamma$) and agro-meteorological (w) stations. Panel (b), different stages of the sampling procedure.

For each sampling planar position three samples were collected respectively in the [0 – 10] cm, [10 – 20] cm and [20 – 30] cm depth horizons for a total number of 48 samples for each set. The gravimetric water content of each soil sample was evaluated after drying the samples at 105°C for about 24 h (Hillel, 1998).

The results of the measurements (Figure 5) show a monotonically increasing trend in water content for increasing depth for the 1 DBI and 3 DAI depth profiles, which is reversed for the 1 DAI validation set. Among the three datasets the deepest soil horizon exhibits a much less pronounced variation in soil water content compared to the superficial layers.

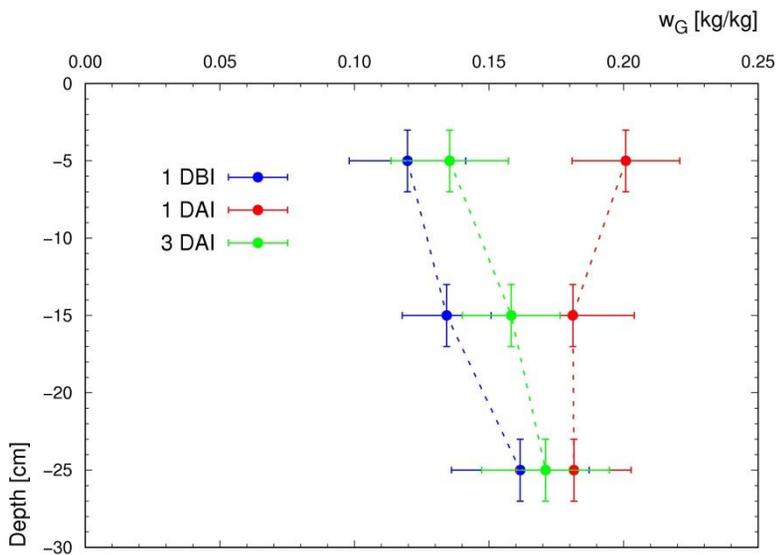

Figure 5. Individual gravimetric soil water contents ($w_G$) for the [0 – 10] cm, [10 – 20] cm and [20 – 30] cm depth horizons are reported in terms of mean and standard deviation of the corresponding 16 samples collected for the three sets of validation measurements performed in the vegetated period: 1 Day Before Irrigation (DBI) (24/07/17), 1 and 3 Days After Irrigation (DAI) (26/07/17 and 28/07/17). Each data point is referred to the median depth, and the vertical error bar represents the 2 cm sampling uncertainty.

Above-ground crop biomass samples were collected by using the destructive sampling method (Catchpole and Wheeler, 1992) at four different maturity stages. The plants, including stems, leaves and fruits, were sampled in different days at the same diurnal time. The water mass of stems plus leaves and of fruits was separately evaluated by drying the samples at 80°C for about 24 h (SERAS, 1994). For each biomass gravimetric sampling campaign, the measured overall water mass per plant (kg/plant) was converted to BWC in mm by adopting the specific density of 3.5 plants/m$^2$. Data on BWC (mm) were linearly fitted to obtain the BWC temporal evolution over the vegetated period (Figure 6).

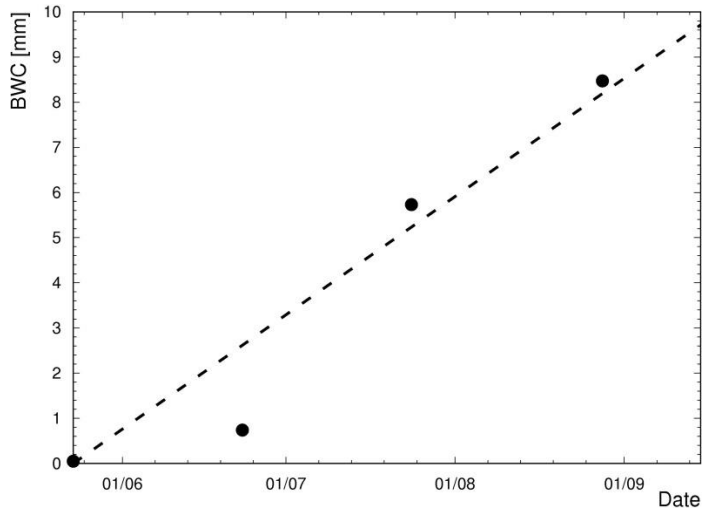

Figure 6. Overall Biomass Water Content (BWC) in mm estimated from destructive gravimetric measurements on stems, leaves and fruits samples collected at four different maturity stages of the tomato crop. Data were fitted according to a linear regression curve with a 0.084 mm/day slope and a 0.921 coefficient of determination.

## 2.3.2. Gamma-ray measurements

The gamma-ray spectrometer detects the photon radiation produced in the decays of natural occurring radionuclides ($^{40}$K, $^{238}$U and $^{232}$Th) and records a list mode output, i.e. a continuous logging of individual photons arrival time and acquisition channel. A dedicated software was developed to post-process the output list mode files in order to (i) generate gamma spectra corresponding to 15 minutes acquisition time, (ii) perform an energy calibration procedure, (iii) remove the spectral background, and (iv) retrieve the net count rate in the main $^{40}$K, $^{214}$Bi ($^{238}$U) and $^{208}$Tl ($^{232}$Th) photopeak energy windows (IAEA, 2003). Average net count rates are 11.51 cps in the $^{40}$K energy window (1.37 – 1.57 MeV), 0.89 cps in the $^{214}$Bi energy window (1.66 – 1.86 MeV) and 2.49 cps in the $^{208}$Tl energy window (2.41 – 2.81 MeV), while the gross counting statistics in the [0.30 – 3.00] MeV range is of about 200 cps. While $^{40}$K and $^{208}$Tl are distributed only in the soil, $^{214}$Bi gamma radiation has an atmospheric component due to the exhaled $^{222}$Rn gas which makes the $^{214}$Bi count rate inadequate for soil water content estimation as it clearly fluctuates in the day-time and in relation to rainfall events (Barbosa et al., 2017). Given also the typically higher net counting statistics in the main photopeak compared to $^{208}$Tl, $^{40}$K is chosen as natural gamma emitter for soil water content assessment purposes.

Thanks to a specifically developed management software, gamma and meteorological data were temporally aligned and merged in a unique database having a 15 minutes temporal resolution and 44 different fields (34 related to gamma measurements and 10 to meteorological measurements). Data were hourly averaged and a statistical fluctuation typically lower than 1% is observed in the net number of events. The global dataset has 20502 entries corresponding to a 5125 hours acquisition time during which both the gamma and agro-meteorological stations were operative, for a 260 GB total amount of raw data.

# 3. Soil water content estimation

## 3.1. Theoretical background

The inverse proportionality between soil moisture and gamma signal is the key point suggesting that gamma-ray spectroscopy can be an operative method for retrieving SWC (Carroll, 1981; Grasty, 1997). (Baldoncini et al., 2018) provides by means of Monte Carlo simulations a proof of concept of the effectiveness and reliability of proximal gamma-ray spectroscopy for the determination of the gravimetric SWC, $w_G$ (kg/kg).

The SWC at time t can be determined by monitoring the counting statistics of a gamma spectrum in the photopeak of energy $E_i$ (Baldoncini et al., 2018) as:

$$w_\gamma^i(t) = \frac{S^{Cal}(E_i)}{S(E_i, t)} \cdot \left[ \Omega(E_i) + w_G^{Cal} \right] - \Omega(E_i)$$

*(1)*

where

$S(E_i, t)$ (cps) is the net count rate in the photopeak of energy $E_i$ at time $t$,

$S^{Cal}(E_i)$ (cps) is the net count rate in the photopeak of energy $E_i$ at the calibration time,

$w_G^{Cal}$ (kg/kg) is the SWC determined on the basis of independent measurements at the calibration time.

The dimensionless factor $\Omega(E_i)$ is defined as:

$$\Omega(E_i) = \Psi(E_i) + \left[ 1 - \Psi(E_i) \right] \cdot f_{H2O}^{struct}$$

*(2)*

where $f_{H2O}^{struct}$ (kg/kg) is the fraction of structural water (i.e. water incorporated in the formation of soil minerals) and $\Psi(E_i)$ corresponds to the ratio between the mass attenuation coefficient of the soil solid portion $\left( \frac{\mu}{\rho} \right)_S$ (cm²/g) and the mass attenuation of water $\left( \frac{\mu}{\rho} \right)_{H2O}$ (cm²/g):

$$\Psi(E_i) = \frac{\left( \frac{\mu}{\rho}(E_i) \right)_S}{\left( \frac{\mu}{\rho}(E_i) \right)_{H2O}}$$

*(3)*

By adopting the specific values referred to the composition of the soil at the experimental site (Baldoncini et al., 2018), Eq. (1) can be numerically written for the [40]K photopeak (1.46 MeV) as:

$$w_{\gamma K}(t) = \frac{S_K^{Cal}}{S_K(t)} \cdot \left[ 0.899 + w_G^{Cal} \right] - 0.899$$

*(4)*

In absence of a detailed mineralogical analysis, a $\Omega = (0.903 \pm 0.011)$ mean value can be employed (Baldoncini et al., 2018). In any case, the uncertainty on the estimated SWC is typically dominated by the systematic uncertainty on the $S_K^{Cal}$ and $w_G^{Cal}$ calibration reference values, implying an almost negligible contribution from the $\Omega$ variability to the ~0.017 kg/kg absolute uncertainty.

In order to extract time-by-time SWC values from proximal gamma-ray spectroscopy measurements it is necessary to take into account a non-constant correction due to the presence of growing vegetation beneath the detector position (Figure 2). Indeed, as the tomato plants mature, the gamma spectrometer receives a progressively reduced gamma flux due to the shielding effect produced by the crop system. The latter can be estimated by modelling stems, leaves and fruits as an equivalent water layer characterized by a given thickness which we express as a BWC in units of mm (Figure 6). In particular, the time dependent correction to be applied to the measured gamma signal $S$ can be expressed as the $\Lambda$ ratio given in:

$$\Lambda_K(BWC)\left[\frac{cps}{cps}\right] = \frac{S_K^{MC}(BWC)}{S_K^{MC}(BWC=0)}$$

*(5)*

It follows that the SWC corrected for the attenuation due to the vegetation $w_{\gamma K}^{\Lambda}$ at time $t$ is given by:

$$w_{\gamma K}^{\Lambda}(t) = \frac{S_K^{Cal} \cdot \Lambda_K(\mathrm{BWC}(t))}{S_K(t)} \cdot \left[ 0.899 + w_G^{Cal} \right] - 0.899$$

*(6)*

With the aim of going after the crop evolution temporal profile, a curve describing the attenuation factor $\Lambda(\mathrm{BWC})$ as function of the BWC was determined by adopting the Monte Carlo simulation method described in (Baldoncini et al., 2018) (Figure 7 panel a). Nine independent simulations were performed by progressively increasing the thickness of the equivalent water layer from BWC = 0 mm up to BWC = 20 mm with steps of 2.5 mm. Simulations were carried out with an initial statistics of $10^9$ emitted photons having 1.46 MeV energy and assigning to the soil source a SWC corresponding to $w_G^{Cal}$. The nine $\Lambda(BWC)$ values were fitted according to a linear regression curve with an intercept fixed by definition to 1 (Figure 7 panel a).

In order to estimate how the attenuation due to vegetation affects the estimation of SWC, we evaluate on the basis of the Monte Carlo results the quantity $\delta$ defined as:

$$\delta(BWC)[\%] = 100 \cdot \frac{w'_{\gamma K} - w_G^{Cal}}{w_G^{Cal}}$$

$$(7)$$

where:

$$w'_{\gamma K} = \frac{S_K^{Cal}}{S_K^{Cal} \cdot \Lambda(BWC)} \cdot \left[ 0.899 + w_G^{Cal} \right] - 0.899 = \frac{1}{\Lambda(BWC)} \cdot \left[ 0.899 + w_G^{Cal} \right] - 0.899$$

$$(8)$$

that corresponds to the SWC that would have been measured without correcting for BWC for fixed $w_G^{Cal}$ in the soil. As shown in Figure 7 panel b, the non-corrected SWC $w'_{\gamma K}$ differs from the $w_G^{Cal}$ = 0.163 kg/kg calibration value (Table 2) by about 70% for a BWC of 7.5 mm, which almost corresponds to the estimated BWC at the tomato harvesting (Figure 6). Therefore, the non-application of a vegetation correction factor $\Lambda(BWC)$ to the measured gamma signal has a large systematic effect on the SWC estimation in proximal gamma-ray spectroscopy (Figure 8).

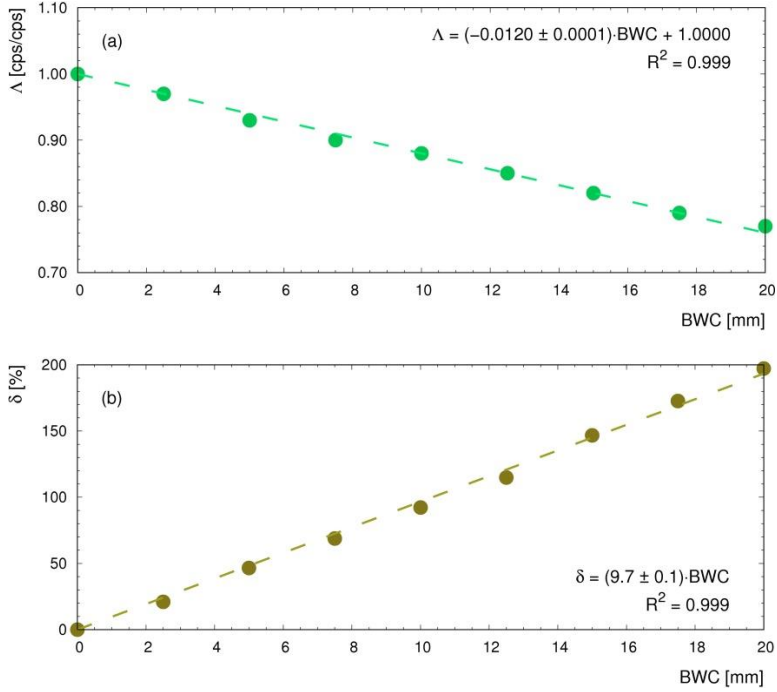

Figure 7. Panel (a), simulated values of $\Lambda$ (Eq. (5)) for the 1.46 MeV $^{40}$K gamma emission energy as function of the BWC fitted with a linear regression curve. Panel (b), percentage overestimation $\delta$ (Eq. (7)) of the SWC as function of the BWC in case no vegetation cover correction is applied.

## 3.2. Experimental results and discussion

The theoretical approach presented in Section 3.1 was applied to the analysis of gamma-ray spectra measured over the entire data taking period; 54% of the data taking was carried out during the vegetated phase. As detailed in Section 2.3.1, a gravimetric sampling campaign was performed with the objective of both calibrating and validating the SWC estimation based on $^{40}$K radiometric data (Table 2). The $w_G$ values referred to [0–10] cm, [10–20] cm, and [20–30] cm were combined with weights respectively equal to 0.79, 0.16, and 0.05, determined on the basis of the depth profile of the expected contribution to the overall gamma signal (Figure 5 of (Strati et al., 2018)). Particular attention was paid in collecting soil samples in different environmental conditions in terms of both temporal proximity to irrigation events and of biomass amount present in the crop. This sampling strategy allowed for testing the reliability of the proximal gamma-ray spectroscopy method as well as for having insights on the bias that the BWC has on the SWC estimation.

Table 2. Results of SWC ($w_G$) for the gravimetric calibration measurement (18 September, one day before a rainfall event) and for four validation measurements. The latter were performed in bare soil condition (21 September, two days after a rainfall event) and during the vegetated period, one day before an irrigation event (24 July), one (26 July) and three days (28 July) after the same event. The $w_G$ values are the weighted average SWC determined from 16 planar sampling points homogeneously distributed within 15 m from the gamma station. For each measurement we report the SWC inferred from proximal gamma-ray spectroscopy measurements without ($w_{\gamma K}$) and with BWC correction ($w_{\gamma K}^{\Lambda}$) together with the corresponding 1σ uncertainty. Δw and Δw$^{\Lambda}$ are the percentage differences between $w_G$ and $w_{\gamma K}$ and between $w_G$ and $w_{\gamma K}^{\Lambda}$, respectively.

| Date | $w_G$ [kg/kg] | $w_{\gamma K}$ [kg/kg] | $w_{\gamma K}^{\Lambda}$ [kg/kg] | Δw [%] | Δw$^{\Lambda}$ [%] |
|---|---|---|---|---|---|
| 18 September | 0.163 ± 0.008 | 0.163 ± 0.017 | 0.163 ± 0.017 | 0 | 0 |
| 21 September | 0.176 ± 0.011 | 0.182 ± 0.017 | 0.182 ± 0.017 | 3 | 3 |
| 24 July | 0.124 ± 0.021 | 0.196 ± 0.017 | 0.126 ± 0.017 | 58 | 2 |
| 26 July | 0.197 ± 0.021 | 0.256 ± 0.017 | 0.181 ± 0.017 | 30 | -8 |
| 28 July | 0.141 ± 0.021 | 0.203 ± 0.017 | 0.133 ± 0.017 | 44 | -6 |

For the validation measurement performed in bare soil condition (21[st] of September) the correction of BWC plays no role, as the value of the attenuation function Λ is identical to 1 for null BWC (see Eq. (5)). The three validation measurements performed on the 24[th], 26[th] and 28[th] of July allow for investigating the effect of the BWC correction as the tomato crop was at about midlife of its growing cycle. If the attenuation due to the presence of BWC is neglected, (see Eq. (4)), the $w_{\gamma K}$ would be affected by a systematic positive bias larger than 30%. By accounting for the attenuation effect of BWC (see Eq. (6)), an excellent agreement between $w_{\gamma K}^{\Lambda}$ and $w_G$ is obtained, with a maximum relative discrepancy below 10% and a 1σ level agreement for all the three validation measurements. Therefore, systematic errors leading to underestimations or overestimation of the SWC are to be excluded also in presence of the tomato crop at the experimental site.

Considering that the BWC changes in time as the tomato crop grows, a non-constant BWC correction was applied to gamma signals over the entire tomato life cycle. The temporal evolution of the attenuation correction factor was estimated on the basis of the BWC temporal growth (see Figure 6).

Figure 8 shows a positive correlation between radiometric inferred SWC and the amount of precipitations that includes both rainfall and irrigation water. Nevertheless, the $w_{\gamma K}^{\Lambda}$ and $w_{\gamma K}$ datasets exhibit significantly different dynamic ranges. In order to have a meaningful interpretation of the SWC variation domain it is necessary to account for the soil hydraulic properties, which are typically used as reference for defining crop water demand. Indeed, the systematic bias obtained by neglecting the BWC correction can lead to non-physical results corresponding to SWC frequently exceeding soil field capacity and sometimes reaching values close to saturation (see Section 2.1), especially when the crop approaches the maturity stage (Figure 8). During the vegetated period, $w_{\gamma K}^{\Lambda}$ values vary from 0.09 kg/kg to 0.21 kg/kg, coherently with the range identified by wilting point (0.07 kg/kg) and field capacity (0.24 kg/kg). Conversely, the $w_{\gamma K}$ values vary from 0.14 kg/kg to 0.33 kg/kg and show a substantial progressive positive drift as the tomato crop matures.

In the perspective of employing proximal gamma-ray spectroscopy for automatic irrigation management, the BWC correction is mandatory for assessing crop water demand and for a sustainable use of water. The method developed in this work for the assessment of the BWC shielding effect relies on the modelling of biomass with equivalent water layers and is in principle independent from the type of crop sowed in the agricultural field. Nonetheless, more field measurements over different vegetation types are desirable in order to confirm the performance of the method, in particular for crops characterized by high water content or in the case of tall trees cultivations.

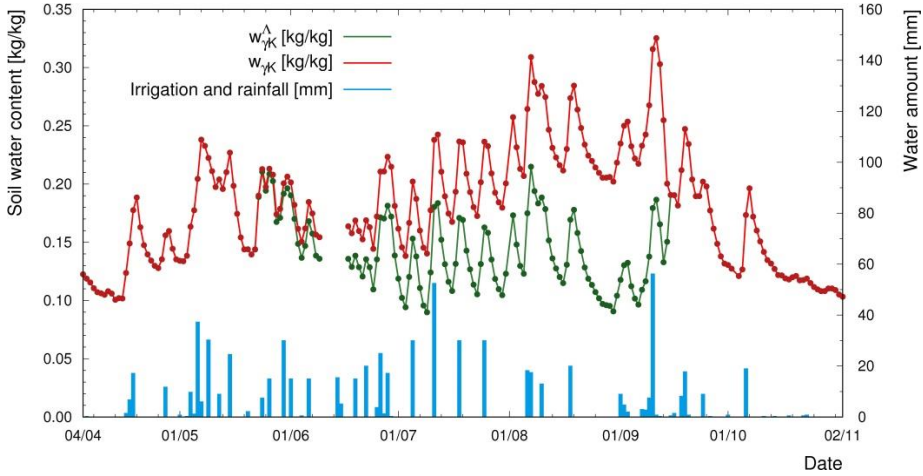

Figure 8. SWC inferred from measured $^{40}$K count rate without ($w_{\gamma K}$, see Eq. (4)) and with ($w^{\Lambda}_{\gamma K}$, see Eq. (6)) BWC correction for the entire data taking period.

## 4. Conclusions

The continuous tracing of soil water content provided by radiometric measurements has high potentialities for a site specific rational irrigation planning aimed at a sustainable use of water. In this study we demonstrate that proximal gamma-ray spectroscopy performed with permanent stations can be considered an effective tool for estimating soil water content for the following reasons: (i) the installation of a proximal gamma-ray spectroscopy station is economically affordable, (ii) the method is able to provide SWC time series with hourly frequency, sensitive to transient soil moisture levels and consistent with soil hydraulic properties, (iii) the method provides a continuous monitoring of SWC with a field scale footprint, filling the gap between punctual and satellite soil moisture measurement techniques, (iv) the results are affected by a ~10% relative uncertainty and are in agreement with independent validation gravimetric measurements on soil samples.

An unbiased quantitative estimate of the gravimetric water content requires a proper correction of the measured gamma signal for the reduction caused by water distributed in the growing vegetation.

We demonstrate that a reliable way to evaluate the shielding due to stems, leaves and fruits is to model biomass as an equivalent water layer which thickness increases during the crop life-cycle. Monte Carlo simulations highlight that gamma-ray measurements are not only extremely sensitive to water in the soil but also to water concentrated in the biomass which acts as a shielding layer sitting on top of the soil gamma source. In particular, the gamma signal is affected by a sizeable reduction on the order of 10% for 10 mm equivalent water thickness, which would translate into a soil water content estimation biased by 90%.

Soil water content inferred from proximal gamma-ray spectroscopy was validated against independent gravimetric measurements. The validation set of measurements performed in bare soil condition provides an excellent result, with a 3% relative deviation of the gamma estimated value from the reference gravimetric one. By applying the BWC correction to gamma measurements acquired during the vegetated period, the systematic positive bias on SWC is prevented and an average relative discrepancy of 4% for the validation measurements is observed. In closing, neglecting the BWC shielding effect would provide overestimated soil water content values implying that proximal gamma-ray spectroscopy would be useless as a monitoring and decision support tool for automatic irrigation scheduling, with negative impacts on crop productivity.

## Acknowledgements

This work was partially founded by the National Institute of Nuclear Physics (INFN) through the ITALian RADioactivity project (ITALRAD) and by the Theoretical Astroparticle Physics (TAsP) research network. The authors would like to acknowledge the support of the Project Agroalimentare Idrointelligente CUP D92I16000030009, of the Geological and Seismic Survey of the Umbria Region (UMBRIARAD), of the University of Ferrara (Fondo di Ateneo per la Ricerca scientifica FAR 2016) and of the MIUR (Ministero dell'Istruzione, dell'Università e della Ricerca) under MIUR-PRIN-2012 project. The authors thank the staff of GeoExplorer Impresa Sociale s.r.l. for their support and Renzo Valloni, Stefano Anconelli, Domenico

Solimando, Marco Bittelli, Vincenzo Guidi, Barbara Fabbri, Enrico Calore, Sebastiano Fabio Schifano, Claudio Pagotto, Ivan Callegari, Gerti Xhixha and Merita Kaçeli Xhixha for their collaboration which made possible the realization of this study. The authors show their gratitude to Giovanni Fiorentini, Miguel García Castaño, Ilaria Marzola, Michele Montuschi and Barbara Ricci for useful comments and discussions. The authors thank the Università degli Studi di Ferrara and INFN-Ferrara for the access to the COKA GPU cluster.